\documentclass{article}
\newcommand{\hide}[1]{}

\usepackage{dsfont}
\usepackage{epsfig}
\usepackage{graphicx}
\usepackage{slashed}
\usepackage{color}

\begin{document}

\markboth{{\bf Fermi and the theory of weak interactions
}} {{ G. Rajasekaran}}

\begin{center}
{\large{\bf FERMI AND THE THEORY OF WEAK INTERACTIONS}}
\vskip0.5cm

{\bf G. Rajasekaran}
\vskip0.35cm
{\it Institute of Mathematical Sciences, Chennai 600113 \\
and Chennai Mathematical Institute, Chennai 603103 \\
e-mail: graj@imsc.res.in}
\vskip0.35cm
\end{center}

\vspace{1.5cm}
\underline{Abstract} 

\vspace{3mm} The history of weak interactions starting with Fermi's
creation of the beta decay theory and culminating in its modern avatar
in the form of the electroweak gauge theory is described. Discoveries of
parity violation, matter-antimatter asymmetry, W and Z bosons and neutrino
mass are highlighted.

\vspace{1cm}

\underline{\bf Introduction}

\vspace {3mm}

Sun gives us light and heat that
makes life possible on the Earth. How do the Sun and stars produce
energy and continue to shine for billions of years? Thermonuclear
fusion is the answer as Eddington proposed in 1920 and Bethe demonstrated
explicitly in 1939. Through a series of nuclear reactions, four
protons (which are Hydrogen nuclei) in the core of the
Sun combine to form a Helium nucleus emitting two positrons and
two neutrinos and releasing 27 MeV of energy:

    $$ p + p + p + p \rightarrow He^4 + e^+ + e^+ + \nu_e +\nu_e + 27 MeV $$

\noindent This can be regarded as the most important reaction for all life,
for without it there can be no life on Earth! \\

The above reaction is caused by one of the basic forces of Nature,
called weak interaction. Beta decays of nuclei and in fact the decays of most of
the elementary particles are now known to be due to weak interaction.\\

Enrico Fermi formulated the theory of weak interactions in 1934 and his
theory has stood the ground very successfully with appropriate amendments
and generalizations and finally served as a core part of the Standard Model
of High Energy Physics, which is now known as the basis of almost ALL of
Physics, except for gravitation.\\ 

In this article we trace the historical
evolution of the theoretical ideas punctuated by the landmark experimental discoveries.
This history can be divided into two parts separated by the year 1972 which
marks the watershed year since the gauge theoretic revolution
that converted Fermi's theory into the modern electroweak theory occured
roughly around that year. Discovery of P and CP violation as well as
the discovery of neutrino oscillation and neutrino mass are weaved into 
this tapestry as integral partds of weak interaction physics and its history.\\

\vspace {5mm}

\underline {\bf Early history: Weak interactions upto 1972}

\vspace {3mm}

The story of weak interactions starts with Henri Becquerel's
discovery of radioactivity in 1896 and its subsequent classification
into alpha, beta and gamma decays of the nucleus by Ernest Rutherford 
and others. But the real
understanding of beta-decay in the sense we know it now came
only after Enrico Fermi invented a physical mechanism for the
beta-decay process in 1934.\\

The basic ingredient for Fermi's theory had been provided by
Wolfgang Pauli.To solve the puzzle of the continuous energy
spectrum of the electrons emitted in the beta-decay of the nuclei,
Pauli had suggested that along with the electron, an almost
massless neutral particle also was emitted. Fermi succeeded in incorporating
Pauli's suggestion and thus was born the theory of weak interactions.
Fermi also named the  particle as neutrino.\\

Drawing an analogy with electromagnetic interaction which at the quantum level is the
emission of a photon by an electron, Fermi pictured the weak
interaction responsible for the beta-decay of the neutron as the
emission of an electron-neutrino pair, the neutron
converting itself into a proton in the process.(Fig 1)

\begin{figure}[h!]
 	\begin{center}\vspace{0.125cm}
 	 \includegraphics[width = 10.0cm]{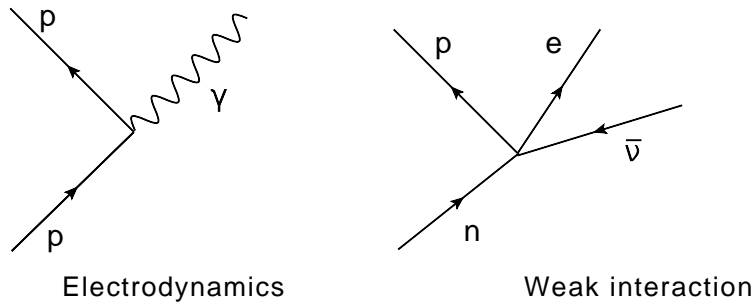}
	\caption{ Fermi's analogy}
	\label{fig1}
 	\end{center}\vspace{0.cm}
\end{figure}

By initiating Quantum Electrodynamics Dirac had laid the foundation for
Quantum Field Theory (QFT) in 1927. Within a few years Fermi made the first
nontrivial application of QFT to weak interactions in which material particles
are created.\\

Either because of the neutrino which most people at that time did not believe in, 
or because of QFT which most people did not understand at that time
or because of both, Fermi's note on beta decay theory was rejected by {\it Nature}
with the comment "it contained speculations too remote from reality to be
of interest to the reader". Fermi then sent the paper to {\it Nuovo Cimento}
which accepted it; another version was published in {\it Zeischrift} [1,2].
(A vivid picture of those times is given in Ref [3]).\\   

In Quantum Electrodynamics (QED) the electromagnetic current $J_E$ of the charged particle
like the electron interacts with A, the electromagnetic vector potential 
which becomes the field operator for the photon: 
 
    $$ L_E = e J_E A. $$

\noindent The symbol $e$ is the numerical value of the electrical charge
of the electron and characterizes the strength of the electromagnetic interaction.
In Fermi's theory of weak interactions, the
weak current of the proton-neutron pair written as $ \bar{p}n $  
interacts with the weak current of the electron-neutrino pair
denoted by $\bar{e}\nu$:

         $$ L_{F} \ = \ \frac{G_F}{\sqrt{2}} ({\bar{p}n \ \bar{e} \nu + \bar{n}p \ \bar{\nu}e}) $$ 

\noindent where the particle symbols $ p, n, e, \nu $ etc represent the corresponding
field operators. Detailed explanation of the language of QFT needed to
understand QED and Fermi's theory is 
given in Appendix 1. 
The strength of the weak interaction is characterized by the Fermi coupling constant
$G_F$ whose value is
         $$ G_F = \frac{10^{-5}}{m_{p}^{2}} $$
where $m_p$ is the proton mass. It is because of the smallness of this number
that this interaction is called "weak" in contrast to the nuclear force
which is "strong".\\ 

The two terms in Fermi's $ L_E $ give rise to the decays of
the neutron and proton:

    $$ n \rightarrow p + e^{-} + \bar{\nu} $$
    $$ p \rightarrow n + e^{+} + \nu $$ 

\noindent Although proton does not decay in free space since it is lighter than
neutron, such decays can occur when nuclei are involved. 
These two terms lead to all nuclear
beta decays.\\

The electric current $J_E$ and the vector potential $A$ are vectors and so
Fermi adopted the vector form for the weak currents also. We suppress the vector 
indices in $J_E$and $A$ and further we do not write Fermi's weak currents 
in the vector form, for simplicity of presentation.\\
  
This theory of
weak interactions proposed by Fermi almost 80 years ago purely on an intuitive basis
stood the test of time inspite of many amendments that were incorporated into
Fermi's theory successfully.
One important amendment came after the discovery of parity violation 
in weak interaction by T D Lee, C N Yang and C S Wu in 1956.(See Appendix 2)
But Fermi's theory survived even
this fundamental revolution and the only modification was to replace
the vector current of Fermi by an equal mixture of vector(V) and
axial vector (A) currents. Vectors and axial vectors behave differently
when we go from left to right-handed coordinate systems and hence the
parity violation.This is the V-A form discovered by 
Sudarshan and Marshak, Feynman and Gell-Mann and Sakurai in 1957.\\

However there was a period of utter confusion before the above correct
form of the interaction was found. As we saw earlier, Fermi based his intuition
on electromagnetism which involves a vector current and we shall see
in the context of later developments how sound this intution proved to be.
Infact this was a master-stroke of Fermi. However, subsequently, 
in an attempt at generalization, Fermi's vector form was replaced by an 
arbitrary combination S,V,T,A,P (scalar, vector, tensor, axial vector
and pseudoscalar) interactions and it led to enormous complication
and confusion in the confrontation of experiments with theory. The
confusion was resolved and the correct V - A form could be found only
because of the additional experimental clues provided by parity violation.\\

During 1947-55, many new particles such as muons, pions, kaons
and hyperons were discovered and all of them were found to decay by
weak interactions. In fact parity revolution itself was triggered
by the famous tau-theta puzzle in the decays of the kaons which
was the culmination of the masterly phase-space plot analysis
of the three-pion decay mode of the kaon by Richard Dalitz. The field of
weak interactions
thus got enriched by a multitude of phenomena, of which nuclear
beta-decay is just one. Weak interaction is indeed a universal
property of all fundamental particles.\\

Remarkably enough, all the weak phenomena, namely the weak decays of
all the particles could be incorporated in a straight-forward
generalization of the original Fermi interaction. This was
achieved by Feynman and Gell-Mann (1957) in the form of the current
x current interaction:

   $$ L_{FG} = \frac {G_F}{2\sqrt{2}} (J_+ J_- + J_- J_+) $$
 
       $$ J_+ = \bar{p}n + \bar{\nu}_e e + \bar{\nu}_{\mu}\mu + ... $$
       $$ J_- = \bar{n}p + \bar{e}\nu_e + \bar{\mu}\nu_{\mu} + ...  $$

\noindent The dots at the end refer to other terms that can be added in order to 
incorporate the weak decays of other particles 
such as the strange particles $\Lambda$, $\Sigma$, and 
$K$. A diagrammatic representation of this is given in Fig 2. The current $J_+$ represents
a neutron turning itself into a proton,
an electron turning itself into a neutrino or a muon turning itsef
into a neutrino - all these transitions result in an increase of electrical charge
by one unit. The weak current $J_+$ is called the charge-raising current; $J_-$
describes the opposite transition and is called the charge-lowering current. \\

\begin{figure}[h!]
 	\begin{center}\vspace{0.125cm}
 	 \includegraphics[width = 12.0cm]{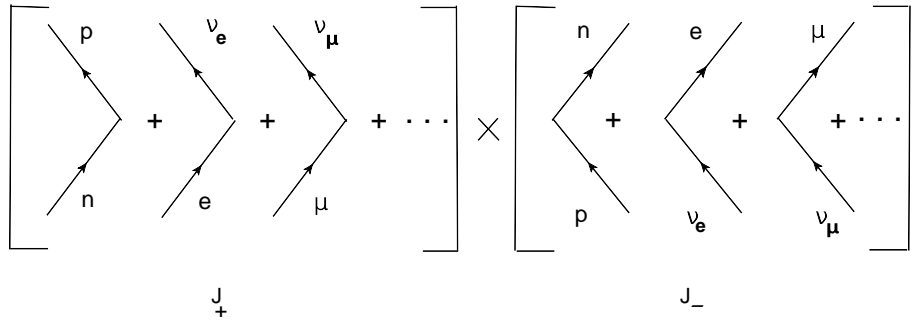}
	\caption{ Current $\times$ current interaction.}
	\label{fig6}
 	\end{center}\vspace{0.cm}
\end{figure}

One can see that Fermi's original form of the interaction describing the beta decay
of the neutron is just one term $ \bar{p}n \ \bar{e}\nu_e $ in the product $ J_+ J_- $.
The decay of the muon and the absorption of the muon by the proton are described by the 
terms $ \bar{\nu}_{\mu}\mu \ \bar{e}\nu_e $ and $ \bar{\nu}_{\mu}\mu \ \bar{n}p $
respectively. These are illustrated in Fig 3. By turning around the line a particle
in the initial state can become an antiparticle in the final state. This can be
understood from Appendix 1 where it is explained that a field operator can annihilate
a particle or create an antiparticle.
This happens for the neutrino in $n$ decay and $\mu$ decay
depicted in Fig 3.\\

\begin{figure}[h!]
 	\begin{center}\vspace{0.125cm}
 	 \includegraphics[width = 12.0cm]{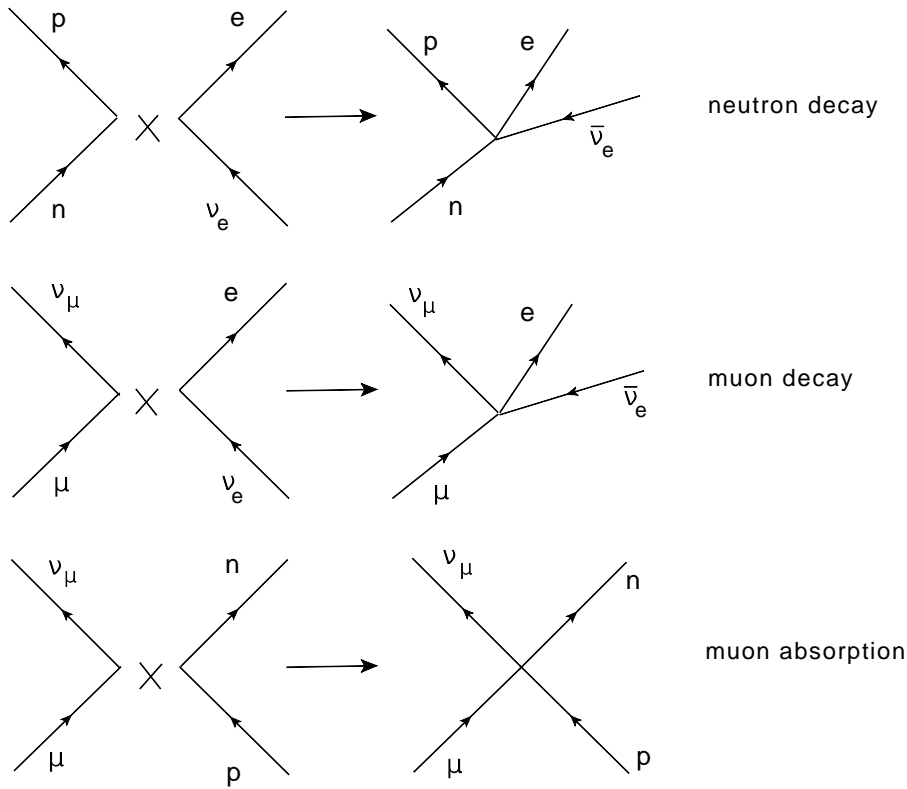}
	\caption{Consequences of current $\times$ current theory. }
	\label{fig7}
 	\end{center}\vspace{0.cm}
\end{figure}

A fundamental experimental discovery - the discovery of CP violation
was made by Cronin and Fitch in 1964, in the weak decays of
neutral kaons. It is this asymmetry in the basic laws of nature 
that is presumed to be responsible for the evolution of the original
matter-antimatter symmetric Universe into the present-day asymmetric Universe that
contains only matter.\\

The story of weak interactions is not complete without due
recognition of the neutrino, especially because of more recent
developments to be described later.\\

Pauli proposed the neutrino in 1930. Although because of the
success of Fermi's theory based on neutrino emission in
explaining quantitatively all the experimental data on nuclear
beta decays, there was hardly any doubt (at least in theorists'
minds) that neutrinos existed, a direct detection of the neutrino
came only in 1956. This achievement was due to Cowan and Reines who
succeeded in detecting the antineutrinos produced from fission
fragments in nuclear reactors.\\

Subsequently, it became possible to detect the neutrinos from
the decays of pions and kaons produced in high-energy
accelerators. It is by using the accelerator-produced neutrinos
that the important experiment proving $\nu_{\mu}$ not to be the same as $\nu_e$
was done.\\

Further, even neutrinos produced by cosmic rays
were detected. The underground laboratory at the deep
mines of the Kolar Gold Fields in South India was one of the first to detect
cosmic-ray produced neutrinos called atmospheric neutrinos. This was in 1965.\\

\vspace {5mm}

\underline {\bf Electroweak theory: Weak interaction after 1972}

\vspace {3mm}

We have already drawn attention to the analogy between weak interactions
and electrodynamics which Fermi exploited in constructing his theory. One
may attribute it to the intuition of Fermi's genius or to just good luck.
Whatever it is, the analogy with electrodynamics that he banked upon
not only made him to choose the correct form of the weak interaction
- the vector form 
in contrast to the scalar or tensor form that were introduced later
but then rejected - but also served as a fruitful analogy in the search 
for a more complete theory of weak interactions. This is what we shall
describe now.\\

In beta decay, Fermi had imagined the $n-p$ line and the $e-\nu$ line
interacting at the same space-time point. But clearly the correspondence
with electrodynamics is greatly enhanced if the two pairs of lines are
separated and an exchange of a quantum W between the $n-p$ and $e-\nu$
lines is inserted.(See Fig 4).

\begin{figure}[h!]
 	\begin{center}\vspace{0.125cm}
 	 \includegraphics[width = 12.0cm]{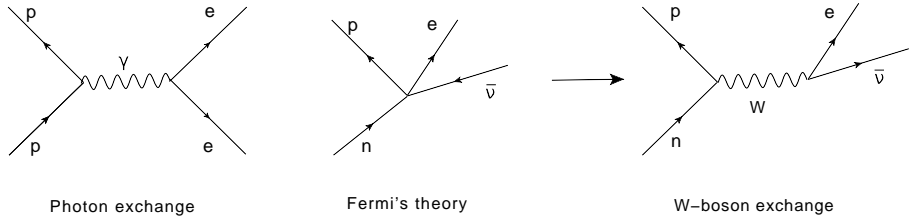}
	\caption{Genesis of W-boson theory.}
	\label{fig8}
 	\end{center}\vspace{0.cm}
\end{figure}

What are the properties of this new quantum or particle?\\
 
(a) W has to be charged, in contrast to the photon, as can be seen by
conserving charge at the two vertices of the W-exchange diagram in Fig 4.
The neutron turns into a proton by emitting a W and so this W should be 
negatively charged.\\

(b) Just like the photon, the W particle also has a spin angular momentum
of one unit. Both photon and W are bosons.\\

(c) In contrast to the photon, the W boson has to be a very massive object.
For, the weak interaction has a short range unlike the infinite-ranged
electromagnetic interaction.\\

In Fermi's theory the coupling constant was $G_F$. In the W-boson theory
we have a coupling constant $g$ at each vertex and so for the same process
$G_F$ is replaced by a factor $g^2$ multiplied by the propagation factor
for the W boson. This propagation factor is $\frac{1}{m^2_W}$, where $m_W$ is the
mass of W, for small energy and momentum transfers relevant in beta decay.
Thus we have the important relationship:

                   $$ G_F = \frac{\sqrt{2} \ g^2}{8 \ m^2_W} $$
                   
By introducing the field $W^+$ and $W^-$ for the positively and negatively charged
W bosons, the current x current form of the weak interaction can be split into
the form
             $$ g(J_+W^+ + J_-W^-). $$
This form is very similar to the electrodynamic interaction $ eJ_EA $ and so
we have achieved a greater degree of symmetry between weak interaction and 
electrodynamics.(See Fig 5.)

\begin{figure}[h!]
 	\begin{center}\vspace{0.125cm}
 	 \includegraphics[width = 12.0cm]{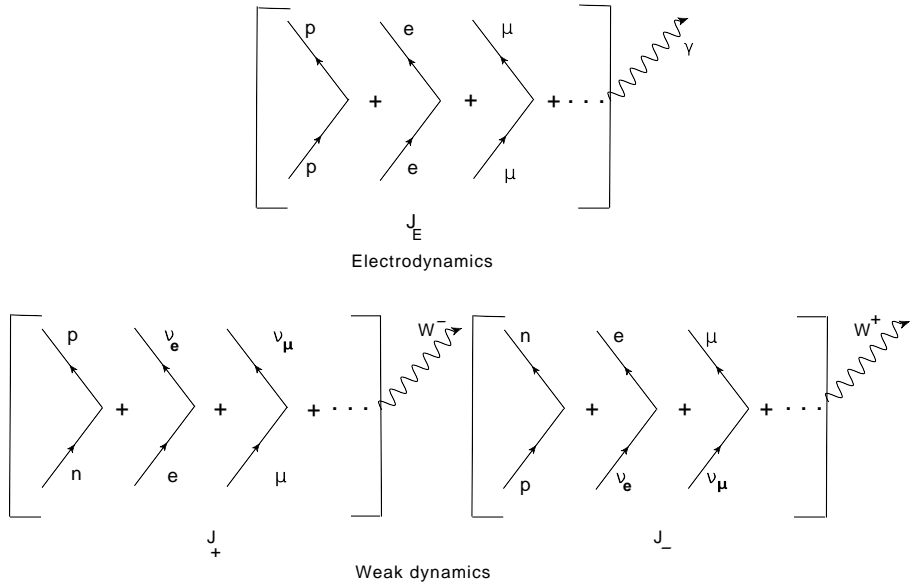}
	\caption{Symmetry between electrodynamics and weak dynamics.}
	\label{fig9}
 	\end{center}\vspace{0.125cm}
\end{figure}

The next step of the argument is to realize that the symmetry between the W boson
form of weak interaction and electrodynamics noted above is only apparent
and does not hold at a deeper level.\\

Conservation of electric charge is a cornerstone of electrodynamics. The
total charge of an isolated system can neither be increased nor decreased
and remains constant. A related question concerns gauge invariance which
simply means that different electromagnetic potentials A lead to the
same physical effects as long as the electric field and the magnetic field
that are obtained by taking space and time derivatives of A are the same.
Are such properties valid for the W boson theory formulated above? The
answer is in the negative.\\

Certain important structural modifications have to be made in the W-boson
theory in order to achieve conservation of the generalized charge involved in 
weak interaction and the corresponding gauge invariance.\\  

The required basic theoretical structure has been known since 1954 when
C N Yang and R Mills introduced nonabelian gauge theory which is
a generalization of electrodynamics. The gauge invariance of electrodynamics
is known as abelian gauge invariance and Yang-Mills theory has 
nonabelian gauge invariance based on a nonabelian Lie group. But many
other ideas had to be discovered before this theory could be tailored
to meet the experimental facts of weak interactions. The final outcome
is the electroweak gauge theory of Glashow, Salam and Weinberg which
is the successor to Fermi's theory.\\

This theory generalizes the concept of charge. The single electric charge
of electrodynamics is replaced in the new theory by four generalized charges.
The current corresponding to each charge interacts with its own boson,
called gauge boson. An essential point of electroweak theory is that the
twin requirements of generalized charge conservation and gauge invariance
force us to combine weak and electromagnetic interactions dynamically 
into a single framework. As a consequence of this unification of weak
and electromagnetic interactions, a new kind of weak interactions is
also generated.\\

The combined interaction in the electroweak theory is
   $$ eJ_{E}A + g (J_+ W^+ + J_-W^-) + g_N J_N Z, $$
where the last term is a new
interaction.There are four generalized charges whose currents $J_E, J_+,J_-$ and $J_N$  
interact with the four gauge bosons:  photon, $W^+,W^- $ and $Z$ respectively. 
Thus electroweak theory introduces 
a symmetry between the photons and the massive weak bosons $ W^+ $, $ W^- $
and $ Z $. Photon becomes a member of a family of four electroweak gauge bosons.\\

The generalization of the concept of charge leads to self-interactions among the
gauge bosons, which are shown in Fig 6. Both a cubic and a qurtic coupling are
present. This is a new feature not present in electrodynamics. The photon interacts
with every electrically charged particle. But the photon itself being uncharged,
does not interact with itself. On the other hand the gauge bosons of Yang-Mills
theory themselves carry the generalized charges and hence they have to interact with 
themselves.

\begin{figure}[h!]
 	\begin{center}\vspace{0.125 cm}
 	 \includegraphics[width = 4.0cm]{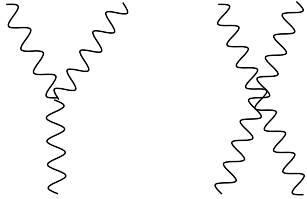}
	\caption{Self-interactions among the gauge bosons.}
	\label{fig10}
 	\end{center}\vspace{0.125cm}
\end{figure}

So we have completed a full circle. We started with Fermi who made his theory of
beta decay by mimicking electrodynamics. We tried to make that copying more and more
perfect. We end up by unifying electrodynamics and beta decay into the same framework.
The myriad electrodynamic and weak decay phenomena are manifestations of one 
electroweak force.
We now discuss some of the simple consequences of electroweak theory\\

\vspace {5mm}

\underline{\bf Neutral-current weak interaction}

\vspace {3mm}
Electroweak theory encompasses not only the known electromagnetic and
weak interactions, but also a new type of weak interaction $g_N J_N Z$.
The current $J_N$ consists of terms such as 
$ \bar{p}p, \bar{n}n, \bar{\nu}\nu $ and $\bar{e}e $
illustrated in Fig 7. In contrast to  $ J_+ $ and $ J_- $ which change electrical charge
the new current describes transitions in which charge does not change
and is called the neutral current. The neutral current interacts with
the neutral weak boson Z with coupling constant $g_N$. 
The neutral current weak interaction would 
lead to neutrino scattering processes in which the neutrino emerges as
a neutrino (with change of energy and direction) rather than getting converted 
into a charged particle such as electron or muon. (see Fig 8.)

\begin{figure}[h!]
 	\begin{center}\vspace{.125cm}
 	 \includegraphics[width = 9.0cm]{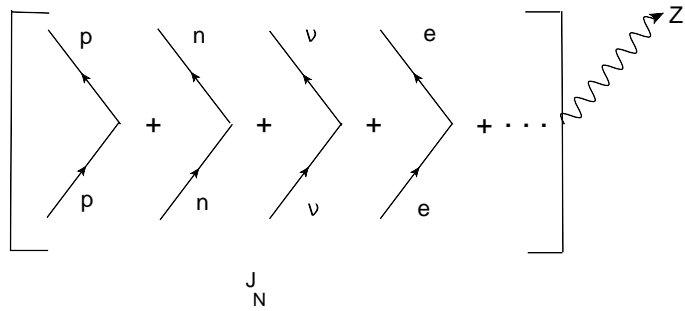}
	\caption{Neutral current interaction.}
	\label{fig11}
 	\end{center}\vspace{0.cm}
\end{figure}

\begin{figure}[h!]
 	\begin{center}\vspace{.125cm}
 	 \includegraphics[width = 6.0cm]{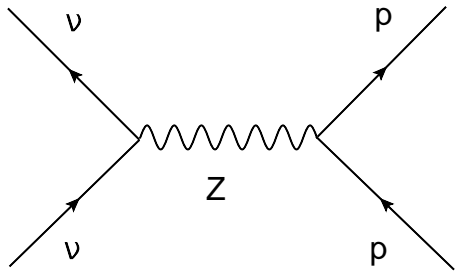}
	\caption{ Neutrino scattering on proton by neutral current interaction.}
	\label{fig12}
 	\end{center}\vspace{0.cm}
\end{figure}

The discovery of the neutral current weak interaction was made in 1973
at CERN, Geneva. This discovery has its own intrinsic importance because
it opened up a whole new class of weak interactions which had remained
undetected in all the 70 years' history of weak interactions. From the
point of view of electroweak theory, it has an added significance since
the neutral-current (NC) interaction acts as a bridge between electrodynamics
and the old charged-current (CC) weak interaction. It is neutral like
electromagnetic current, but involves a massive boson Z like the W involved
in the CC interaction. Hence its discovery with properties identical to
those predicted by the electroweak theory was the first great triumph
of the theory.\\

In the history of weak interaction physics the discovery of the V-A structure
of the charged current was an important milestone. What is the structure
of the neutral current? It is not V-A and so not all of weak interaction 
is described by V-A theory! The relative amount of V and A in
neutral current is specified by an important parameter of the electroweak 
theory called the weak mixing angle $ \theta_W $ and it
has been determined experimentally: 
        $$ sin \ \theta_W = 0.23. $$ \\ 

\vspace {5mm}

\underline {\bf Discovery of W and Z}

\vspace{3mm}

An immediate consequence of the dynamical connection between weak and
electromagnetic interactions is that their coupling constants are related:
       $$ e = g \  sin \ \theta_W, $$
       $$ g_N = \frac{g}{cos \ \theta_W} $$
and the masses of W and Z are also related:
        $$ m_W = m_Z \ cos \ \theta_W. $$
The relationship between $G_F$ and $g^2$ derived earlier 
        $$ G_F = \frac{\sqrt{2} \ g^2}{8 \  m^2_W} $$
now becomes
        $$ G_F = \frac{\sqrt{2} \ e^2}{8  \\ \sin^{2} \ \theta_W \  m^2_W} $$. 

This allows us to calculate the masses of W and Z from the known values
of $G_F$, $e$ and $sin \ \theta_W$. We get
       $$ m_W =  80 \ GeV.   $$

       $$ m_Z =  91 \ GeV.   $$
The discovery of W and Z with these masses at CERN in 1982 was
the second great triumph of electroweak theory.\\

The inverse relationship between $G_F$ and and $m^2_W$ given above helps us
to answer the question: why is weak interaction weak? It is because
the masses of $ m_W $ (and $m_Z$) are so large. $G_F$ is the effective weak
coupling constant at low energies. Once the energy becomes high enough to
produce a real W boson, weak interaction attains its real strength $g$
which is comparable to $e$, the strength of the electromagnetic interaction.  

\vspace {5mm}

\underline {\bf Spontaneous breaking of symmetry and the Higgs boson}

\vspace {3mm}

An essential ingredient of the electroweak theory described sofar is
spontaneous breakdown of symmetry, also known as Higgs mechanism.
The gauge invariance or gauge symmetry of Yang-Mills theory would
lead to massless gauge bosons exactly as the gauge invariance of
electrodynamics requires massless photon. But we need massive 
gauge boson to describe the short-ranged weak interaction. How
is this problem solved in electroweak theory? It is solved by the
spontaneous breakdown of symmetry engineered by the celebrated
Higgs mechanism which keeps photon massless while raising the masses
of W and Z to the finite values discussed above. For more on the
Higgs mechanism see Ref [5]. \\

An important byproduct of the Higgs mechanism is the existence of
a massive spin zero boson, called the Higgs boson. High energy
physicists searching for it in all the earlier particle accelerators
and colliders could not find it. Finally in 2012, the Higgs boson
with a mass of 125 GeV was discovered at the Large Hadron Collider 
at CERN. Higgs boson remained as the only missing piece in the
electroweak theory and with its discovery electroweak theory is
fully established.\\

Electroweak theory and quantum chromodynamics (which is the theory
of strong interactions) have become the twin pillars of the Standard Model
of High Energy Physics [5].\\

\vspace {5mm}

\underline {\bf Renormalizability and precision tests}

\vspace {3mm}

It had been known for a long time that Fermi's original form of the
weak interaction (in which four fermionic fields meet at a single point
in a contact interaction) can only be regarded as an effective potential
to be used in the lowest order approximation to a perturbative calculation.
Any attempt to improve the accuracy of the result by calculating the
next order in perturbation leads to infinity which does not make any sense.\\

Construction of a dynamical theory of weak interaction free from this defect
was one of the fundamental problems of high energy physics. Electroweak theory
solved this problem. Stated in technical language electroweak theory is
renormalizable in the same sense as Quantum Electrodynamics (QED) is. The
renormalizability of electroweak theory was proved by Gerard t'Hooft
and Martinus Veltman in 1972.\\

Renormalizability of the electroweak theory elevated it to a theory
whose precision now rivals that of QED, which is considered as the most
precise theory constructed by man. Electroweak theory came out with
flying colours in all the precision tests performed through a series of
experiments at the Large Electron Positron Collider (LEP) at CERN in the
1990's.\\    

\vspace {5mm
}
\underline {\bf Transition to the quark era}

\vspace{3mm}

Sofar in the article, we have used proton and neutron to describe the
weak interaction. But we now know that proton and neutron are composed of
quarks $u$ and $d$. Proton is made up of $uud$ and neutron is made up of
$ddu$. At the fundamental level weak interaction acts on the quarks.  The
currents therefore must be rewritten in terms of the quark fields. In
the beta decay of the neutron it is one of the two $d$ quarks that decays
into $u$ as shown in Fig 9. The other quarks play only spectator role.

\begin{figure}[h!]
 	\begin{center}\vspace{.125cm}
 	 \includegraphics[width = 6.0cm]{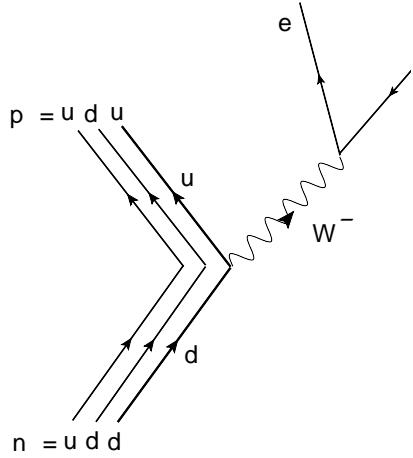}
	\caption{Beta decay of the neutron in terms of quarks.}
	\label{fig13}
 	\end{center}\vspace{0.0cm}
\end{figure}

There exist 6 types of quarks arranged in the form of 3 doublets
        $$ (u,d) \  (c,s) \  (t,b). $$
All composite particles formed out of these quarks are called hadrons.
Our familiar proton and neutron are hadrons and many more hadrons are known.
Electron and neutrino have remained elementary on par with quarks upto
the present. These are called leptons and again 6 types of leptons are
known to exist:
        $$ (\nu_1,e) \  (\nu_2,\mu) \  (\nu_3,\tau). $$

The electroweak interaction in terms of the quarks and leptons is given
by
       $$ L_{EW} = eJ_EA + g(J_+W^+ + J_-W^-) +g_NJ_NZ. $$

\noindent The elecromagnetic and neutral currents will contain terms like
       $$ \bar{u}u, \ \bar{d}d, \ \bar{e}e \ ... $$
while the charged currents  that describe the transitions from one type
of quark to another or from one type of lepton to another (as illustrated
in Fig 10) are given by
 
$$ J_+ = \bar{u}d + \bar{c}s +\bar{t}b +\bar{\nu}_1e +\bar{\nu}_2\mu +\bar{\nu}_3\tau$$
$$ J_- =\bar{d}u+\bar{s}c+\bar{b}t+\bar{e}\nu_1+\bar{\mu}\nu_2+\bar{\tau}\nu_3 $$

\begin{figure}[h!]
 	\begin{center}\vspace{.125cm}
 	 \includegraphics[width = 8.0cm]{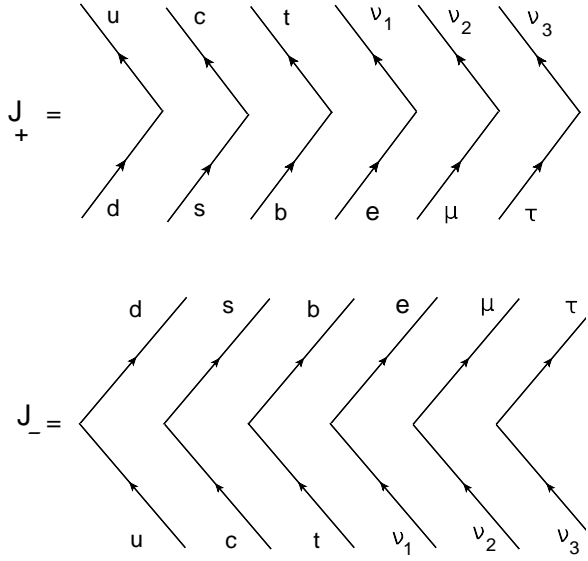}
	\caption{ The currents in terms of quarks and leptons.}
	\label{fig14}
 	\end{center}\vspace{0.cm}
\end{figure}

The electric charge of the "up-type" quarks $(u,c,t)$ is $+\frac{2}{3}$ while
that of the "down-type" $(d,s,b)$ is $-\frac{1}{3}$. All the transitions between the
up and down type of quarks indicated by the above expression or Fig 10 have
a change of charge by one unit and are of the same sign, exactly like the transitions
between the charged leptons $(e,\mu,\tau)$ and the neutrinos.\\

Actually we have to change the down quarks $(d, s, b)$ occuring in $J_+$ and $J_-$ by their
linear superpositions $(d^{\prime}, s^{\prime}, b^{\prime})$ defined as follows.
Intruducing the notation $q_i (i = 1,2,3)$ for $(d,s,b)$, the superposed quarks
are given by
          $$ q_{i}^{\prime} = \sum_j U_{ij}q_j $$
where U is a $3\times3$ unitary matrix:
          $$ U^{\dagger}U = 1.$$

\noindent This U is called the CKM matrix and its discovery by Cabibbo,
Kobayashi and Maskawa is an important chapter in the history of weak interactions.
(For more details on this part of history, see Ref [6]). These superpositions are
natural consequences in electroweak theory and
allow the heavier quarks to decay into all the lighter quarks. The matrix $U$ is
parametrized by three angles and one phase that is responsible for the
CP violation discovered by Cronin and Fitch.\\

A similar unitary mixing matrix $V$ called PMNS matrix (named after Pontecorvo,
Maki, Nakagawa and Sakata) is used in the leptonic part of the currents also
and this is what leads to neutrino oscillation and the discovery of neutrino mass.
(For more on this, see Ref [7].)\\   

It is important to note that all the 6 quarks and 6 leptons are equally
fundamental and all were presumably created in equal numbers in the Big Bang
and it is the weak interaction that caused all the heavier
particles to decay into the lighter ones $u, d, e, \nu$ that make up the 
fermionic or matter-component of the present-day
Universe. \\

\vspace{5mm}

\underline {\bf Discovery of neutrino mass}

\vspace {3mm}

In his original paper in 1934 Fermi had already come to the conclusion 
that the mass of the neutrino was either zero or very small as compared
to that of the electron (0.5 MeV), by comparing the energy distribution
of the electrons emitted in beta decay near their end-point energy with
what was available experimentally. When the electroweak theory was constructed,
massless neutrinos had a natural place in the theory. So neutrinos were considered
massless.\\

However neutrino oscillations were discovered in 1998 by the SuperKamioka
group in Japan and this implied mass differences among the three neutrinos
and hence neutrinos have mass. This discovery was made in the study of 
atmospheric neutrinos which had been first detected in India in 1965, as
already mentioned.\\

Actually, indications for neutrino oscillations and neutrino mass came
first as early as 1970 from the pioneering solar neutrino experiments
of Davis etal in USA which were later corroborated by many other solar
neutrino experiments. But the clinching evidence that solved the solar
neutrino problem in terms of neutrino oscillations had to wait until
2002 when the Sudbury Neutrino Observatory (SNO) could detect the
solar neutrinos through both the neutral current as well as the
charged current weak interactions. In the introduction we had mentioned
that it was the thermonuclear fusion reactions caused by weak interaction
that powered the Sun and the stars. The experimental proof of this too
came from the SNO experiment.(See ref [8,9] for a more complete description) \\ 

Although neutrinos are now known to have mass from the existence of
neutrino oscillations, we do not know the values of the masses since
only differences in neutrino mass-squares can be determined from oscillation
phenomena. The two differences between the mass-squares of the three types
of neutrinos have been determined to be 

   $$ m^{2}_{2} - m^{2}_{1} = 7 \times 10^{-5} eV^2 $$

   $$ |m^{2}_{3} - m^{2}_{2}| = 2 \times 10^{-3} eV^2 $$

Going back to Fermi's original comment, since more accurate measurements
on the end-point energy disribution in nuclear beta are possible now
compared to 1934, what can be said? Progressively the upper limits 
on the neutrino mass determined by this method have come down and
the present upper limit from Tritium beta decay is 2.2 eV, which is 
indeed very small. All the three neutrino masses are clustered around a value lower than
this upper limit, with tiny mass-differences between them.\\    

It is important to point out that even 80 years after its birth, 
the fundamental nature of the neutrino is still not known, namely
whether neutrino is its own antiparticle or not. If it is its own
antiparticle it is called Majorana particle; otherwise it is a Dirac particle
just like the other fermions such as electron or quark. This question
can be answered only by the "neutrinoless double beta decay experiment"
which is therefore the most important experiment in all of neutrino physics
(See Ref [10] for an account of the Majorana problem).\\

Neutrino physics is now recognized as one of the most important frontiers
in high energy physics and it is vigorously pursued in many underground
laboratories around the world. The India-based Neutrino Observatory (INO)
that is coming up in South India will be one such [11].\\

As already mentioned, electroweak theory implies massless neutrinos
in a natural way. How is the theory to be extended to incorporate
nonzero neutrino masses? Only Future will tell.

\vspace {5mm}

\underline {\bf Epilogue}

\vspace {3mm}

We have seen the vast range of phenomena covered by weak interactions:
beta decay of nuclei, thermonuclear fusion reactions in the Sun and
stars, decays of most of the elementary particles of Nature and removal
of antiparticles in the Universe through CP violation. We have touched
on the brief history of the important theoretical and experimental
discoveries. The milestones in this history are listed in Appendix 3.\\

Fermi created beta decay theory which was the starting point of all
that followed, using the nascent Quantum Field Theory which was perhaps
understood by very few physicists at that time. He did this at a time
when nuclei were not understood and so nuclear physics did not even
exist - not to speak of particle physics (now called high energy physics)
which was born only much later. No wonder Fermi responded that it is
beta decay theory when asked what he regarded as his most important
contribution. There is no doubt that it is not only Fermi's most important
contribution but it is one of the most important contributions made by anybody
in that Foundational Epoch of Modern Physics.\\  

\vspace{5mm}

\underline {\bf Acknowledgements} 

\vspace{3mm}

The author thanks Prof N Mukunda for 
inviting him to write this article. He thanks Dr Saurabh Gupta for help
in drawing the large number of figures that the article contains.

\vspace {5mm}

\underline {\bf References}

\vspace{3mm}

\noindent 1. E. Fermi, Attempt at a Theory of $\beta$-rays, Il Nuovo Cimento, Vol 11, p 1 (1934); Zeitschrift fur Physik, Vol 88, p 161 (1934) \\
2. For Fermi's article in English, see The Development of Weak Interaction Theory,
   Ed: P K Kabir, Gordon and Breach, Science Publishers, (1963). This volume
   contains most of the original papers in weak interaction physics upto 1962.\\
3. A Pais, Inward Bound, Oxford University Press, New York, 1986, p 417-418.\\
4. S Chaturvedi and Shyamal Biswas, Fermi-Dirac Statistics, Resonance, Jan 2014, p 45.\\
5. G Rajasekaran, Standard Model, Higgs Boson and What Next?, Resonance, Oct 2012, p 956.\\
6. G Rajasekaran, Cabibbo Angle and Universality of Weak Interactions, Physics News, 
   July 2011, p 21.\\
7. G Rajasekaran, An angle to tackle the neutrinos, Current Science,103,No 6 (2012)\\
8. G Rajasekaran, Hans Bethe, the Sun and the Neutrinos, Resonance, Oct 2005, p 49\\
9. D Indumathi, MVN Murthy and G Rajasekaran, Neutrinos and how to catch them, Physics
   Education, Vol 23, No 2, Aug 2006\\
10. G Rajasekaran, Are neutrinos Majorana particles? arXiv:0803.4387[physics]\\
11. More information on INO can be obtained from www.ino.tifr.res.in\\ 

\vspace{5mm}

\underline {\bf Appendix 1: Quantum Field Theory (QFT)}

\vspace {3mm} 
Quantum Field Theory created by Dirac and used by Fermi to describe
weak interactions remains to this day as the correct basic language
to understand ALL high energy physics. Here we give an elementary
account of its symbolism and interpretation.\\

We start with Quantum Electrodynamics. The electromagnetic force between two
charged particles such as proton and electron can be represented by the
diagram in Fig 11. The wavy line denotes the photon which is the quantum of
the electromagnetic field and it is exchanged between the proton and the electron.
It is this exchange that leads to the electromagnetic force between the charged
particles. This is the quantum version of the classical picture where
the proton is considered to produce the electromagnetic field which then
influences the electron placed in the field.

\begin{figure}[h!]
 	\begin{center}\vspace{.125cm}
 	 \includegraphics[width = 5.5cm]{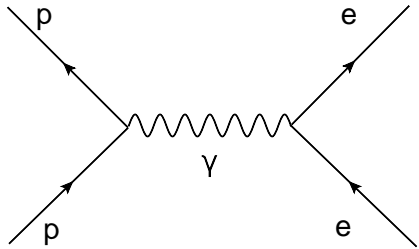}
	\caption{Exchange of photon. }
	\label{fig2}
 	\end{center}\vspace{0.125cm}
\end{figure}

In QFT, the range of a force is inversely proportional to the mass of the quantum 
that is exchanged. Since photon mass is zero, the electromagnetic force mediated
by the exchange of photons is of infinite range.\\

We shall make a rather liberal use of pictorial representations of 
interactions and processes such as in Fig 11. These are called Feynman diagrams, after
Feynman whose use of such diagrams in an intuitive interpretation of
complex calculations in QFT was an important step in the elucidation
of the fundamental processes of Nature.\\

One can split up the above diagram into two parts, one describing the
emission of the photon by the proton and the other describing the 
absorption of the photon by the electron. So the
basic QED interaction in symbolic form is written as  $ eJ_E A $.
Here  $ A $ denotes the quantum field operator for the photon and
$ J_E $ denotes the electromagnetic current of the charged particles.\\

The photon field operator A is actually the quantum version of the
vector potential of classical electrodynamics (from which the electric
field and magnetic field can be obtained by taking suitable space
and time derivatives). So all the fields become quantum operators
whose main property is that they can either create or annihilate 
particles. In this case, A can create or annihilate a photon, thus
explaining the emission or absorption of a photon.\\

Further the current $ J_E $ also is composed of field operators, but
field operators of the charged particles like proton and electron.
In QFT every particle is a quantum of the corresponding
field; electron is a quantum of the electron field. 
Denoting the proton and electron field operators by $p$ and $e$ respectively,
we write $ J_E $ symbolically as a sum of terms $\bar{p}p, \bar{e}e$ etc.\\
  
In general, a field operator such as $p$ can either annihilate a particle or create
the corresponding antiparticle while a field operator with a
bar above such as $\bar{p}$ can either annihilate an antiparticle or create a particle.
The antielectron is the positron. In general the antiparticle is
different from particle. But for photon antiparticle is not different; 
antiphoton is the same as
photon and so A annihilates or creates the photon only.
Hence the interaction described by the combined operator $ \bar{e}e A $
actually describes the 8 processes given in Fig 12.\\ These describe the emission
or absorption of a photon by an electron or positron and also the annihilation
or creation of an electron-positron pair. Similar processes exist with
proton and antiproton.

\begin{figure}[h!]
 	\begin{center}\vspace{.125cm}
 	 \includegraphics[width = 12.0cm]{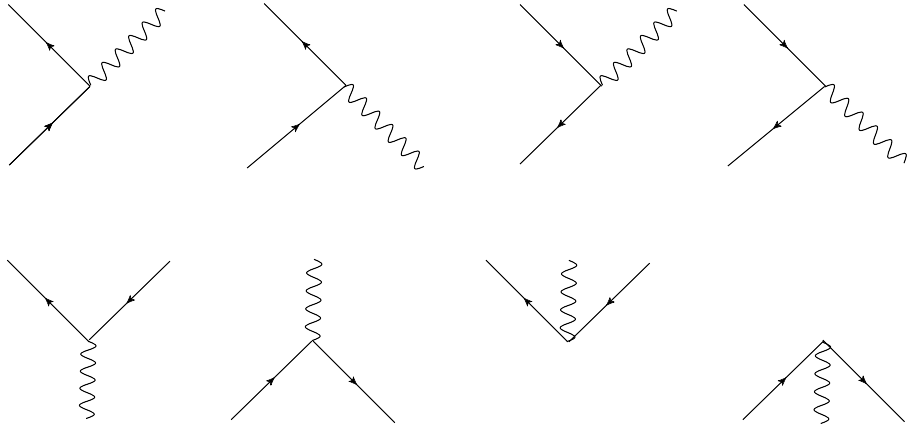}
	\caption{ The basic QED processes.}
	\label{fig3}
 	\end{center}\vspace{0.125cm}
\end{figure}

In all Feynman diagrams time increases vertically upwards. Particles and
and antiparticles are distinguished by having their arrows in the upward
and downward directions respectively.\\     
 
All the processes in Fig 12 are virtual processes to be used as basic elements
in buiding the diagrams of real physical processes, the simplest of them
being the scattering of an electron by a proton depicted in Fig 11.\\

This is the basic symbolism of QFT that has to be kept in mind when
we describe Fermi's theory and its subsequent development.\\ 

In Fermi's interaction $ L_F $ (given in the main text) the  terms
$ \bar{p}n \ \bar{e}\nu $ and $ \bar{n}p \ \bar{\nu}e $ contain the
field operators $ p, n, \nu $ and $ e $ and their barred counterparts.
These create or annihilate the corresponding particle or antiparticle
as explained above. Hence $ L_F $ leads to many related weak processes apart
from the decay of proton and neutron.\\

Particles or quanta of fields come in two varieties, bosons and fermions.
Bosons are particles with integral spins, photon with spin 1 being their main 
representative and they follow Bose-Einstein statistics. Fermions have
half-integral spins, electron with spin $\frac{1}{2}$ being a fermionic example 
and they follow Fermi-Dirac statistics [4]. Since in Fermi's $ L_F $
four fermionic lines meet at a point, it is also called the four-fermion 
interaction.

\vspace {5mm}

\underline {\bf Appendix 2: Violation of left-right symmetry and CP}

\vspace{3mm}

Left-right symmetry is also called reflection symmetry or parity symmetry 
and is denoted by P. 
Madam Wu's famous experiment which established P violation was done
using the beta decay of $Co^{60}$ nuclei. She aligned the spins of Cobalt 
nuclei by an external magnetic field produced by a circulating current
and counted the number of beta electrons emitted in all directions. She found
more beta electrons emitted in the direction of the magnetic field as
compared to the opposite direction. This was the discovery of P violation.\\ 

In the right-handed coordinate system, the directions of the x, y and z 
axes are such that, if we imagine a screw (actually a right-handed screw
which is what we normally use) being rotated from x to y, the screw will advance
along z. The left-handed coordinate system is obtained by mirror-reflection.
(See Fig 13).

\begin{figure}[h!]
 	\begin{center}\vspace{.125cm}
 	 \includegraphics[width = 12.0cm]{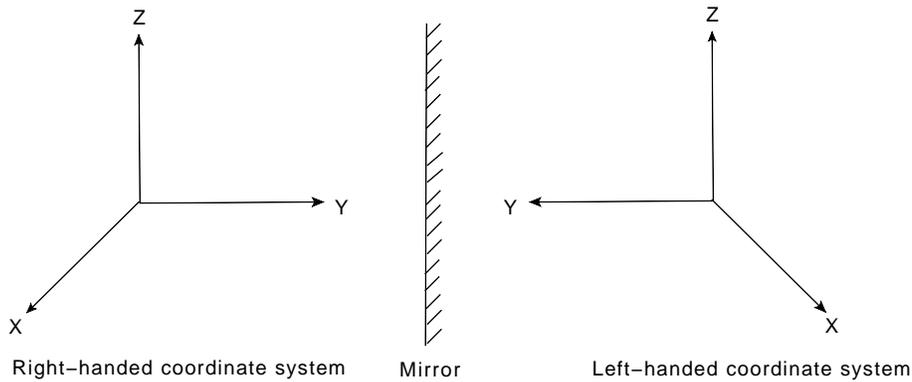}
	\caption{ Reflection in a mirror.}
	\label{fig4}
 	\end{center}\vspace{0.125cm}
\end{figure}

Can the laws of physics distinguish between the two coordinate systems?
Except for the weak interaction, all other laws of physics are symmetric
under mirror reflection and hence cannot be used to distinguish between
the left and right coordinate systems.\\

The significance of this left-right symmetry, as well as its violation can be appreciated better, if
we think of the following attempt at intergalactic communication.\\

Suppose we want to communicate with somebody in a distant galaxy through
radio waves. How do we define a right-handed coordinate system for him?
Screws will not help here since we do not know whether they use a right-
handed screw or a left-handed one in that galaxy! We can use any of the 
laws of physics for this purpose. If none of the laws distinguishes
between the two coordinate systems, we will never be able to convey a
definition of the right-handed coordinate system to a being in a distant
part of the Universe.\\

However, thanks to weak interactions, this can be done. The following
instruction can be conveyed: " Take $Co^{60}$ nuclei and arrange a 
sufficient number of electrons to go around these nuclei, thus forming
an electric current. If a rotation from the x axis to the y axis is
in the direction of the circulating electrons, then the z axis is the
direction in which more electrons are emitted." This would define the
right-handed coordinate system for our friend in the distant galaxy.(See Fig 14).
Thus weak interaction allows us to define a right-handed coordinate
system by using natural physical laws.

\begin{figure}[h!]
 	\begin{center}\vspace{.125cm}
 	 \includegraphics[width = 11.0cm]{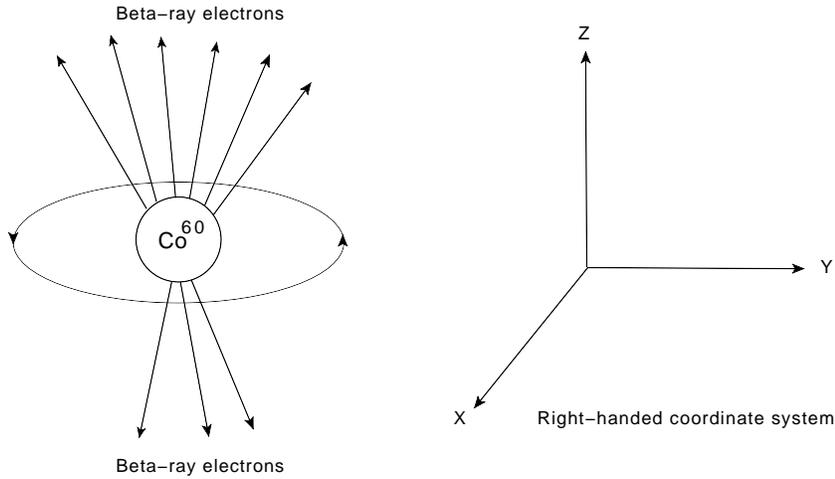}
	\caption{ Parity violation in the beta decay of Co$^{60}$.}
	\label{fig5}
 	\end{center}\vspace{0.cm}
\end{figure}

A word of caution, however. We have to make sure that the planet inhabited
by our friend is made of matter and not of antimatter. If it is made of
antimatter,he would really take nuclei of anti-Co nuclei and electric current
made of positrons and would end up with a left-handed coordinate system
by following our instructions! \\

What is stated in the last para above is the result of CP symmetry, C standing
for particle-antiparticle conjugation. In other words, weak interactions
violate P symmetry and C symmetry. But if both C and P are applied together
weak interactions remain invariant. It was thought until 1964 that CP symmetry 
remains intact in weak interactions. We now know that even this is not correct,
as a consequence of the discovery of CP violation by Cronin and Fitch.
 
 \vspace{10mm}

\underline {\bf Appendix 3: Milestones in the history of weak interactions}

\vspace{3mm}

\noindent 1896  Discovery of radioactivity (Becquerel) \\
1930  Birth of neutrino (Pauli) \\
1934  Theory of beta decay (Fermi) \\
1939  Theory of thermonuclear fusion in the Sun (Bethe and Wesszacker) \\
1954  Nonabelian gauge theory (Yang and Mills)\\
1956  Discovery of parity violation (Lee, Yang and Wu)\\
1956  Detection of the neutrino (Cowan and Reines)\\
1957  Discovery of V-A (Sudarshan, Marshak and others)\\
1957  Current $\times$ current formulation (Feynman and Gell-mann)\\
1961  SU(2) $\times$ U(1) as the electroweak group (Glashow)\\
1964  Discovery of CP violation (Cronin and Fitch)\\
1964  Abelian Higgs mechanism (Higgs and others)\\
1967  Nonabelian Higgs-Kibble mechanism (Kibble) \\
1967  Electroweak theory (Salam and Weinberg)\\
1972  Renormalizability of EW theory (t'Hooft and Veltman)\\
1973  Discovery of neutral current (55 physicists at CERN)\\
1973  CKM phase for CP violation (Kobayashi and Maskawa)\\
1982  Discovery of W and Z (Rubbia and Van der Meer)\\
1992  Precision tests of EW theory (International Collaboration at CERN)\\
1998  Discovery of neutrino mass (Davis, Koshiba and others)\\
2002  Experimental proof of thermonuclear fusion in the Sun (SNO)\\
2007  Verification of CKM theory of CP violation (KEK, Stanford)\\
2012  Discovery of Higgs boson (ATLAS and CMS Collaborations, CERN)

\end{document}